\documentclass[prb,twocolumn,showpacs,amsmath,amssymb,superscriptaddress,floatfix]{revtex4}

\usepackage{graphicx}
\usepackage{dcolumn}
\usepackage{bm}
\usepackage{epsfig}

\begin{document}

\title{Inverse quantum spin Hall effect generated by spin pumping from precessing magnetization into a graphene-based two-dimensional topological insulator}

\author{Son-Hsien Chen}
\email{d92222006@ntu.edu.tw}
\affiliation{Department of Physics and Astronomy, University
of Delaware, Newark, DE 19716-2570, USA}
\affiliation{Department of Physics, National Taiwan University, Taipei 10617, Taiwan}
\author{Branislav K. Nikoli\' c}
\affiliation{Department of Physics and Astronomy, University
of Delaware, Newark, DE 19716-2570, USA}
\author{Ching-Ray Chang}
\email{crchang@phys.ntu.edu.tw}
\affiliation{Department of Physics, National Taiwan University, Taipei 10617, Taiwan}

\begin{abstract}
We propose a multiterminal nanostructure for electrical probing of the quantum spin Hall effect (QSHE) in two-dimensional (2D) topological insulators. The device consists of a ferromagnetic (FM) island  with precessing magnetization that  {\em pumps} (in the absence of any bias voltage) pure spin current symmetrically into the left and right adjacent 2D TIs  modeled as  graphene nanoribbons with the intrinsic spin-orbit (SO) coupling. In the reference frame rotating with magnetization, the device is mapped onto a DC circuit with twice as many terminals whose effectively half-metallic ferromagnetic electrodes are biased by the frequency $\hbar \omega/e$ of the microwave radiation driving the magnetization precession at the ferromagnetic resonance conditions. The QSH regime of the six-terminal \mbox{TI$|$FM$|$TI} nanodevice, attached to two longitudinal and four transverse  normal metal electrodes, is characterized by the SO-coupling-induced energy gap, chiral spin-filtered edge states within finite length TI regions, and  quantized spin Hall conductance when longitudinal bias voltage is applied,  despite the presence of the FM island. The same unbiased device, but with  precessing magnetization of the central FM island, blocks completely pumping of total spin and charge currents into the longitudinal electrodes while generating DC transverse charge Hall currents. Although these transverse charge currents are {\em not} quantized, their induction together with {\em zero}  longitudinal charge current is a unique electrical response of TIs to pumped pure spin current that cannot be mimicked by \mbox{SO-coupled} but topologically trivial systems. In the corresponding two-terminal inhomogeneous \mbox{TI$|$FM$|$TI} nanostructures, we image spatial profiles of local spin and charge currents within TIs which illustrate transport confined to chiral spin-filtered edges states while revealing concomitantly the existence of interfacial spin and charge currents flowing around \mbox{TI$|$FM} interfaces and penetrating into the bulk of TIs over some short distance.
\end{abstract}


\maketitle

\section{Introduction} \label{sec:introduction}

The recent theoretical predictions~\cite{Kane2005,Bernevig2006,Murakami2008} for the quantum spin Hall effect (QSHE) have attracted considerable attention by both basic and applied research communities. In the conventional SHE,~\cite{Nagaosa2008} which manifests in multiterminal devices~\cite{Nikoli'c2006} as pure spin current $I^{S_z}_{\rm T}$ in the transverse electrodes driven by longitudinal unpolarized charge current in the presence of intrinsic (due to band structure) or extrinsic (due to impurities) spin-orbit (SO) couplings, the spin Hall conductance $G_{\rm SH}= I^{S_z}_{\rm T}/V$  can acquire any value ($V$ is the bias voltage applied between the longitudinal electrodes).~\cite{Nagaosa2008,Nikoli'c2006} Conversely, $G_{\rm SH}=2 \times e/4\pi$ becomes quantized in four-terminal devices that exhibit QSHE.~\cite{Kane2005}

The QSHE introduces an example of a new quantum state of matter---the so-called {\em topological insulator} (TI)~\cite{Murakami2008,Kane2005a,Xu2006a,Qi2008} in two dimensions---which is a band insulator with a usual energy gap in the bulk, but which also accommodates {\em gapless} spin-polarized quantum states confined around the sample edges. Unlike closely related quantum Hall insulators,~\cite{Jain2007} where the bulk energy gap and edge states appear due to an external magnetic field, TIs are time-reversal invariant systems whose intrinsic SO coupling  opens a bulk gap while generating the Kramers doublet of edge states. These edge states force electrons of opposite spin  to flow in opposite directions along the edges of the sample. Since time-reversal invariance ensures the crossing of the energy levels of such peculiar chiral and spin-filtered (or ``helical''~\cite{Bernevig2006,Murakami2008}) edge states at special points in the Brillouin zone, the spectrum of a TI cannot be adiabatically deformed into topologically trivial insulator without such states.~\cite{Kane2005a,Xu2006a,Qi2008}

The recent experiments~\cite{Konig2007,Konig2008} on HgTe quantum wells have confirmed some of the anticipated signatures~\cite{Bernevig2006} of QSHE in line with transport taking place through helical edge states, such as: ({\em i}) reduction of the two-terminal charge conductance; ({\em ii}) its independence on the sample width; and ({\em iii}) sensitivity to an external magnetic field that destroys the TI phase. Nevertheless, this is still perceived as an indirect and incomplete detection because it did not confirm that conducting electrons along the edge were spin-polarized. The very recent nonlocal transport measurements on multiterminal HgTe microstructures in the QSH regime suggest that charge transport in these devices occurs through extended helical edge channels~\cite{Roth2009} since their results can be explained via simple application of the Landauer-B\"{u}ttiker scattering theory of multiprobe conductance to TIs attached to several metallic electrodes.~\cite{Buttiker2009}

Undoubtedly, the most convincing evidence for QSHE would be to measure quantized $G_{\rm SH}$, as the counterpart of the quantized charge Hall conductance~\cite{Jain2007} in the integer QHE exhibited by two-dimensional electron gases (2DEGs).  However, this is quite difficult since pure spin currents~\cite{Nagaosa2008,Nikoli'c2006} typically have to be converted into some electrical signal to be observed.~\cite{Saitoh2006,Valenzuela2006,Seki2008} Thus, {\em the key issue for unambiguous QSHE detection}, as well as for {\em the very definition} of what constitutes direct experimental manifestation of 2D TIs, is to design~\cite{Qi2008} devices where {\em electrical} quantities can be measured that are directly related to helical edge state transport and the corresponding quantization of $G_{\rm SH}$.

The experiments~\cite{Saitoh2006,Valenzuela2006,Seki2008} on the inverse SHE---where injection of pure spin current into a device with SO couplings leads to deflection of both spins in the same direction and corresponding Hall voltage between lateral sample boundaries or charge current in the transverse electrodes---provide guidance for QSHE probing via conventional electrical measurements. The devices constructed for these experiments are very flexible and allow for multifarious convincing tests confirming SHE physics.~\cite{Seki2008} In particular,  one of the inverse SHE experiments  has injected pure spin current, generated by spin pumping~\cite{Tserkovnyak2005}  from a ferromagnetic (FM) layer with precessing magnetization driven by RF radiation at the ferromagnetic resonance (FMR) conditions, into a metal with SO couplings to observe the transverse Hall voltage.~\cite{Saitoh2006}

\begin{figure}
\centerline{\psfig{file=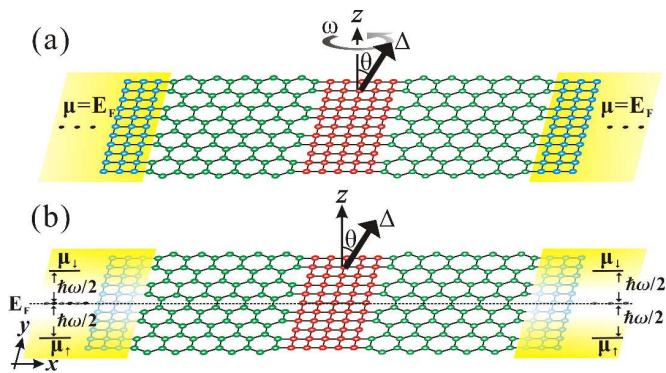,scale=0.32,angle=0}}
\caption{(Color online) (a) The proposed TI$|$FM$|$TI heterostructure consists of a central ferromagnetic island with precessing magnetization which pumps pure spin currents toward the left and the right 2D topological insulators modeled as graphene nanoribbons with intrinsic SO coupling. In the rotating frame in which magnetization is static, device (a) is mapped onto a four-terminal DC circuit in panel (b), whose effectively half-metallic ferromagnetic electrodes have electrochemical potential shifted by $\pm \hbar \omega/2$ with respect to the Fermi level $E_F$ of unbiased normal metal electrodes in the laboratory frame.}
\label{fig:setup}
\end{figure}

Here we propose a spin pumping-based nanostructure, illustrated in Fig.~\ref{fig:setup}(a), where a FM island with precessing magnetization {\em pumps}, in the absence of any  applied bias voltage, pure spin current into two adjacent graphene nanoribbons (GNR) with intrinsic SO coupling that act as the simplest model of 2D TI.~\cite{Kane2005,Kane2005a} This setup has an advantage~\cite{Tserkovnyak2005} over other possible sources of pure spin currents because it evades the conductivity mismatch~\cite{Nagaosa2008} between metallic injector and TI that would play a detrimental role when spin injection is driven by a bias voltage.~\cite{Valenzuela2006,Seki2008} The three-layer central sample is attached to two metallic electrodes, and we also analyze six-terminal setups where additional four transverse electrodes, two per each GNR region, cover a portion of its top and bottom edge.

The nonequilibrium Green function (NEGF) picture of pumping,~\cite{Chen2009,Hattori2007} which was utilized~\cite{Chen2009} to explain very recent experiments~\cite{Moriyama2008} on spin pumping across a band insulator within a magnetic tunnel junction,  converts the complicated time-dependent problem posed by the device in Fig.~\ref{fig:setup}(a) into a four-terminal DC circuit in Fig.~\ref{fig:setup}(b) in the rotating reference frame.  This picture also motivates the usage of a symmetric TI$|$FM$|$TI nanostructure since in asymmetric devices the central FM island would pump~\cite{Chen2009} concomitantly a small charge current into the TI regions (with quadratic frequency dependence as opposed to dominant pumped spin current which is linear in frequency in the adiabatic limit). Within this framework we obtain the following principal results: ({\em i}) in the two-terminal device in Fig.~\ref{fig:setup}(a), pumping generates both spin and charge {\em local} currents inside the TI, but only  {\em total} spin current is non-zero (Figs.~\ref{fig:localpump} and ~\ref{fig:totalpump}); ({\em ii}) imaging of local quantum transport in this two-terminal device also demonstrates the existence of interfacial spin and charge currents at the \mbox{TI$|$FM} boundary, which are able to penetrate into the bulk of TIs over some short distance (Fig.~\ref{fig:localpump}); and ({\em iii}) in the corresponding six-terminal device whose TI regions are brought into the QSH regime, pumped  total spin and charge currents vanish in the longitudinal leads, while non-zero charge currents $I_p$ emerge in the four transverse electrodes ({\em p}=3--6) as the manifestations of the inverse QSHE (Fig.~\ref{fig:sixterminal}). The charge conductances $G_{\rm T}=eI_p/\hbar\omega$, however, are not quantized.

It is worth recalling that 2D TIs, also denoted as QSH insulators, were initially studied as isolated infinite homogeneous systems.~\cite{Murakami2008} On the other hand, experiments require to embed such materials into circuits where they will be attached to multiple metallic electrodes serving as current or voltage probes.~\cite{Roth2009} For example, the very recent analysis has predicted highly unusual spin dynamics at the \mbox{TI$|$normal-metal}~\cite{Yokoyama2009} and \mbox{TI$|$FM}~\cite{Maciejko2009} interfaces. This, together with our findings of interfacial currents around \mbox{TI$|$FM} junction  highlights the need to understand operation of inhomogeneous nanostructures, with multiple \mbox{TI$|$metal} interfaces and helical edge states of {\em finite extent},~\cite{Roth2009} as a prerequisite for the design of anticipated spintronic devices~\cite{Maciejko2009} exploiting TIs. Similar issues, albeit without spin dynamics, were encountered in the well-known experiments~\cite{Haug1989} on inhomogeneous (e.g., containing a potential barrier) multiterminal quantum Hall bridges where a simple picture of edge transport becomes insufficient and one has to take into account  trajectory network~\cite{Haug1989,Gagel1996} connecting edge channels across the bulk of the device.~\cite{Jain2007}

The paper is organized as follows. In Sec.~\ref{sec:negf}, we discuss the effective Hamiltonian of the device and NEGF approach to computation of pumped spin and charge currents in the picture of the rotating reference frame. The images of {\em local} pumped spin and charge currents are shown in Sec.~\ref{sec:two_terminal} for unbiased two-terminal device, while  {\em total} quantized spin Hall currents in biased and pumped charge currents in the transverse electrodes of unbiased six-terminal devices are studied in Sec.~\ref{sec:six_terminal}. Section~\ref{sec:discussion} is devoted to explaining the origin of non-quantized transverse charge currents and zero total spin and charge current in the longitudinal electrodes of our inverse QSHE device using the picture of possible  Feynman paths~\cite{Buttiker2009} of spin-polarized injected and collected electrons within the 12-terminal DC device in the rotating frame. We also support the conjectured network of such paths using spatial profiles of local currents computed for the 12-terminal DC device. We conclude in Sec.~\ref{sec:conclusions}.

\section{NEGF approach to spin pumping in multiterminal QSH systems} \label{sec:negf}

The central \mbox{TI$|$FM$|$TI} region of the nanostructure in Fig.~\ref{fig:setup} can be described by the effective single $\pi$-orbital tight-binding Hamiltonian in the laboratory frame:
\begin{eqnarray}\label{eq:hlab}
\lefteqn{\hat{H}_{\rm lab}(t) =   \sum_i  \hat{c}_{i}^\dag \left( \varepsilon_{i} - \frac{\Delta_{i}}{2} \mathbf{m}_{i}(t) \cdot \hat{\bm \sigma} \right)  \hat{c}_{i} } \nonumber \\
&&{} - \gamma \sum_{\langle ij \rangle} \hat{c}_{i}^\dag \hat{c}_{j}
+ \frac{2i}{\sqrt{3}} \gamma_{\rm SO} \sum_{\langle \langle ij \rangle \rangle} \hat{c}_i^\dagger \hat{\bm \sigma} \cdot ({\bf d}_{kj} \times {\bf d}_{ik})\hat{c}_j.
\end{eqnarray}
 Here $\hat{c}^\dagger_i = (\hat{c}^\dagger_{i\uparrow}, \hat{c}^\dagger_{i\downarrow})$ is the vector of spin-dependent electron creation operators and $\hat{\bm \sigma}=(\hat{\sigma}_x,\hat{\sigma}_y,\hat{\sigma}_z)$ is the vector of the Pauli matrices. For simplicity, the usual (spin-independent) nearest-neighbor hopping is assumed to be the same $\gamma_{sl}=\gamma_c=\gamma_{\rm GNR}=\gamma$ for the square lattice $\gamma_{sl}$ of FM island and the semi-infinite leads, as well as for the hopping $\gamma_c$ that couples different regions of the device. On the honeycomb lattice of TI regions, \mbox{$\gamma_{\rm GNR} \simeq 2.7$ eV} can reproduce~\cite{Areshkin2009a} {\em ab initio} computed band structure very close to the charge neutral Dirac point ($E_F=0$).

 The coupling of itinerant electrons to collective magnetic dynamics is described through the material-dependent exchange potential $\Delta_{i}$, which is non-zero only in the FM island. The on-site potential $\varepsilon_{i}$ can accommodate disorder,~\cite{Sheng2005b,Qiao2008} external electric field, or band bottom alignment. The Hamiltonian~(\ref{eq:hlab}) is time-dependent since the spatially uniform unit vector ${\bf m}(t)$ along the local magnetization direction is precessing  steadily around the $z$-axis with a constant cone angle $\theta$ and frequency $f=\omega/2\pi$. The central \mbox{TI$|$FM$|$TI} region is attached to two or six semi-infinite ideal (spin and charge interaction free) electrodes, modeled on the same square lattice as the FM island, which terminate in macroscopic reservoirs held at the same electrochemical potential $\mu_p=E_F$.

The third sum in Eq.~(\ref{eq:hlab}) is non-zero only in the GNR regions where it introduces the intrinsic SO coupling compatible with the symmetries of the honeycomb lattice. The SO coupling, which is responsible for the band gap~\cite{Kane2005}  $\Delta_{\rm SO}=6\sqrt{3}\gamma_{\rm SO}$, acts as spin-dependent next-nearest neighbor hopping where $i$ and $j$ are two next-nearest neighbor sites, $k$ is the only common nearest neighbor of $i$ and $j$, and ${\bf d}_{ik}$ is a vector pointing from  $k$ to $i$.

Among several proposals for QSHE,~\cite{Murakami2008,Konig2008} graphene with $\gamma_{\rm SO} \neq 0$ stands out as the {\em simplest} model employed to introduce QSHE phenomenology,~\cite{Kane2005} $\mathbb{Z}_2$ classification of TIs,~\cite{Kane2005a} and manifestations of QSHE in realistic multiterminal devices,~\cite{Sheng2005b,Qiao2008} as well as to analyze disorder~\cite{Xu2006a,Sheng2005b,Qiao2008} and  interaction~\cite{Xu2006a,Zarea2008} effects on TIs. However, recent scrutiny finding minuscule \mbox{$\Delta_{\rm SO} \simeq 1$ $\mu$V} through density functional theory (DFT)  calculations~\cite{Min2006} would push the requirement \mbox{$k_BT \ll \Delta_{\rm SO}$} for the observation of QSHE in graphene toward unrealistically low temperatures~\cite{Konig2008} \mbox{$T \ll 10^{-2}$ K}. However, these calculations are not fully first-principles and all-electron---a direct {\em all-electron} (i.e., no pseudopotential) DFT approach~\cite{Boettger2007,Gmitra2009} has estimated $\Delta_{\rm SO} \simeq 0.6$ K, which makes the device in Fig.~\ref{fig:setup} experimentally realizable. The physical explanation of an apparent controversy between extremely small intrinsic SO coupling found in Refs.~\onlinecite{Min2006} and non-negligible values computed in  Refs.~\onlinecite{Boettger2007,Gmitra2009} is in the fact that  96\% of the intrinsic SO splitting in graphene originates from the usually neglected $d$ and higher  orbitals.~\cite{Gmitra2009}

\begin{figure}
\centerline{\psfig{file=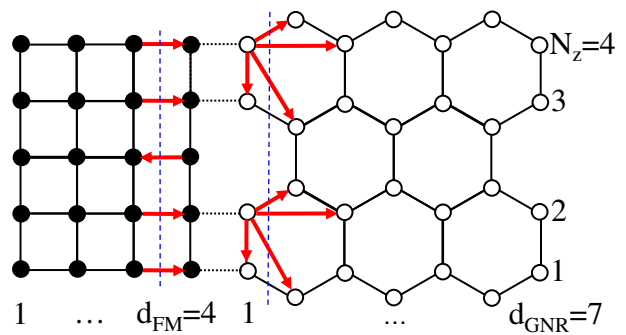,scale=0.5,angle=0}}
\caption{(Color online) Illustration of the definition of bond spin and charge currents between nearest neighbor sites of the square lattice of FM island and nearest neighbor or next-nearest neighbor sites of the honeycomb lattice on which the 2D TI Hamiltonian is defined. The vertical dashed lines show cross sections within FM
or the TI region where the bond currents passing through them are summed up to get the total currents in Fig.~\ref{fig:totalpump} or conductances in Fig.~\ref{fig:sixterminal}.}
\label{fig:curdef}
\end{figure}

We select the following device parameters in the analysis and Figures below: $\gamma_{\rm SO}=0.03 \gamma$, $\Delta=0.5\gamma$, $E_F=10^{-6}\gamma$, $f=2$ GHz and $\theta = 10^\circ$. The geometry and size of zigzag GNR (ZGNR) and FM regions of the device shown in Fig.~\ref{fig:setup} is characterized by: \mbox{$N_z$-ZGNR} is composed of $N_z$ zigzag chains so that its average width is $W=a\sqrt{3}(N_z-1)/2$ (\mbox{$a=2.46$ \AA}   is the lattice spacing of the honeycomb lattice); $d_{\rm GNR}$ is the number of atoms along the zigzag chain defining its length $L=(d_{\rm GNR}-1)a/2$; and $d_{\rm FM}$ is the thickness of the FM island. For illustration of $N_z$, $d_{\rm GNR}$, and $d_{\rm FM}$ parameters see Fig.~\ref{fig:curdef}.

Note that the selected value~\cite{Kane2005} for $\gamma_{\rm SO}=0.03 \gamma$ is much larger than the one that would be fitted to first-principles calculations.~\cite{Boettger2007,Gmitra2009} The reason for this choice (in fact, even larger values have been employed in recent studies~\cite{Sheng2005b,Qiao2008,Zarea2008}) is the usage of the effective tight-binding Hamiltonian~(\ref{eq:hlab}) which in the case of vastly different energy scales $\gamma_{\rm SO} \ll \gamma$ would make quantum transport calculations insensitive to the presence of $\gamma_{\rm SO}$. On the other hand, this does not affect any conclusions about quantized spin transport properties of a graphene-based model of a TI since they do not depend on the particular value of $\gamma_{\rm SO}$ and simply require to perform measurements at temperatures \mbox{$k_BT \ll \Delta_{\rm SO}$} where the band gap of the TI is visible while its Fermi energy is within such gap.

 Although widely-used scattering theory~\cite{Tserkovnyak2005} of adiabatic spin pumping by \mbox{FM$|$normal-metal} (FM$|$NM) interfaces, typically combined with the spin-diffusion equation and magnetoelectric circuit theory, cannot handle nanostructures containing insulators of band (due to spin accumulation not being well-defined in them) or topological type (due to necessity to take into account details of one-dimensional transport through helical edge states), the NEGF approach to spin pumping~\cite{Chen2009,Hattori2007}  can describe both cases by taking the microscopic Hamiltonian~(\ref{eq:hlab}) as an input. The unitary transformation of Eq.~(\ref{eq:hlab}) via $\hat{U}=e^{i \omega \hat{\sigma}_z t/2}$ [for ${\bf m}(t)$ precessing counterclockwise] leads to a time-independent Hamiltonian in the rotating frame:~\cite{Chen2009,Hattori2007}
\begin{equation}\label{eq:hrot}
\hat{H}_{\rm rot}  =  \hat{U} \hat{H}_{\rm lab}(t) \hat{U}^{\dagger} - i \hbar \hat{U} \frac{\partial}{\partial t} \hat{U}^\dagger = \hat{H}_{\rm lab}(0)  - \frac{\hbar \omega}{2} \hat{\sigma}_z.
\end{equation}
The Zeeman term $\hbar \omega\hat{\sigma}_z/2$, which emerges uniformly in the sample and the NM electrodes, will spin-split the bands of the NM electrodes, thereby providing a rotating frame picture of pumping based on the four-terminal DC device in Fig.~\ref{fig:setup}(b). This term breaks time-reversal invariance (while conserving spin $S_z$), but for typical FMR frequencies~\cite{Moriyama2008} of the order of $f \sim$ 1 GHz, it is smaller than~\cite{Boettger2007,Gmitra2009} $\Delta_{\rm SO} \simeq 50$ $\mu$V.

\begin{figure*}
\centerline{\hspace{0.07in}\psfig{file=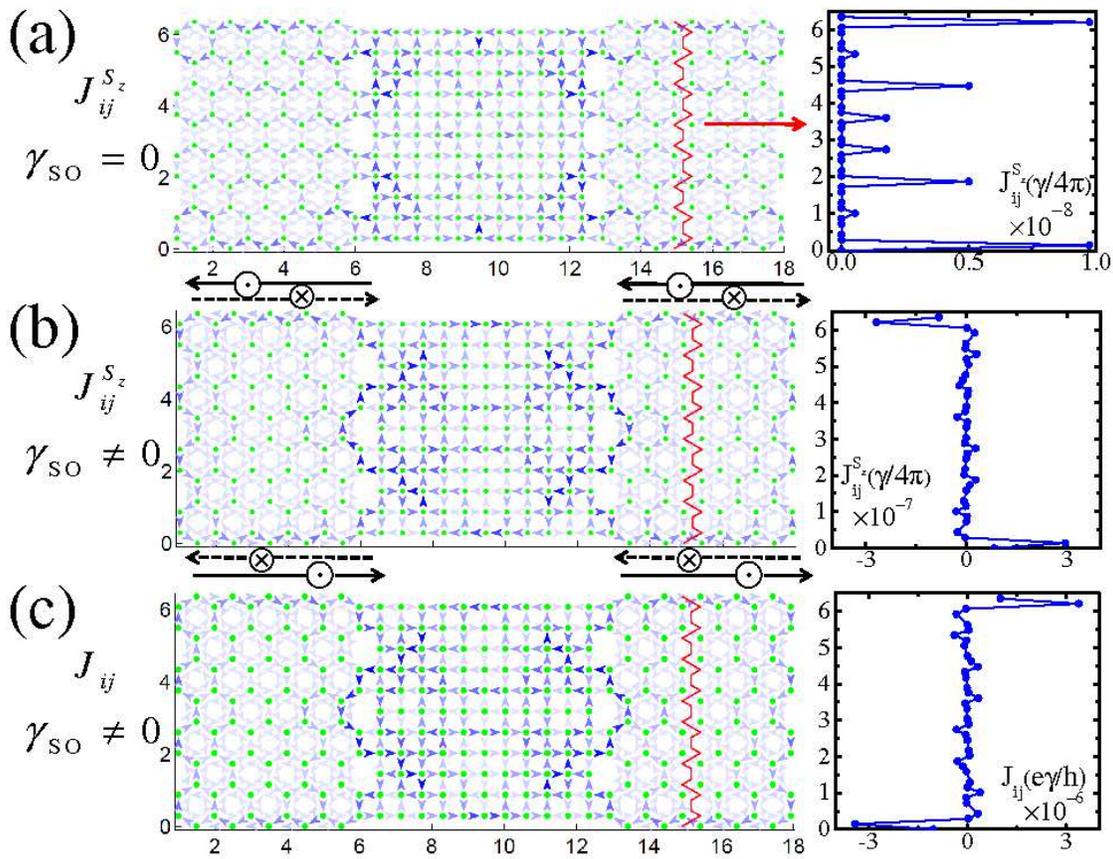,scale=0.6,angle=-90}}
\caption{(Color online) Spatial profile of nearest-neighbor and next-nearest-neighbor bond spin currents (darker arrow means larger current) pumped from the middle FM region of length $d_{\rm FM}=11$ into the left and right \mbox{8-ZGNRs} of length $d_{\rm GNR}=11$ with: (a) zero  or (b) non-zero intrinsic SO coupling for which GNRs act as 2D TI. (c) Spatial profile of bond charge currents corresponding to (b). Right panels show bond current distribution over a selected transverse cross section (vertical twisty line). The FM magnetization is precessing with frequency \mbox{$\hbar\omega/e \approx 8.3$ $\mu$V} and cone angle \mbox{$\theta = 10^\circ$}.}
\label{fig:localpump}
\end{figure*}

The basic transport quantity for the DC circuit in Fig.~\ref{fig:setup}(b) is the spin-resolved bond charge current~\cite{Nikoli'c2006,Onoda2005a} carrying \mbox{spin-$\sigma$} electrons from site $i$ to site $j$
\begin{equation}\label{eq:jlab}
J_{ij}^\sigma = \frac{e}{h} \int\limits_{-\infty}^{\infty} dE\, [\gamma_{ij} \bar{G}^{<,\sigma \sigma}_{j i}(E) - \gamma_{ji}\bar{G}^{<,\sigma\sigma}_{ij}(E)].
\end{equation}
This is obtained in terms of the lesser Green function~\cite{Haug2007} in the rotating frame~\cite{Chen2009,Hattori2007} $\bar{G}^<(E)$. Unlike  $G^<(t,t^\prime)$ in the laboratory frame, $\bar{G}^<$ depends on only one time variable $\tau=t-t^\prime$ (or energy $E$ after the time difference $\tau$ is Fourier transformed~\cite{Nikoli'c2006}). This yields spin
\begin{equation}
J^S_{ij}=\frac{\hbar}{2e}\left(J^{\uparrow}_{ij}-J^{\downarrow}_{ij}\right),
\end{equation}
and charge
\begin{equation}
J_{ij}=J^{\uparrow}_{ij}+J^{\downarrow}_{ij},
\end{equation}
bond currents flowing between nearest neighbor or next-nearest neighbor sites $i$ and $j$ if they are connected by hopping $\gamma_{ij} \neq 0$. The definition of bond currents is illustrated in Fig.~\ref{fig:curdef}, both for the square lattice of FM and the honeycomb lattice of the chosen 2D TI model.

The rotating frame four-terminal device in Fig.~\ref{fig:setup}(b), or its twelve-terminal counterpart  originating  from a six-terminal QSH bridge [illustrated by insets in Fig.~\ref{fig:sixterminal}], guides us in constructing the NEGF equations for the description of currents flowing between their electrodes. The electrodes in the rotating frame are labeled by ($p,\sigma$) [$p=L,R$ and $\sigma=\uparrow,\downarrow$] and they are biased by the electrochemical potential difference $\mu_p^\downarrow - \mu_{p^\prime}^\uparrow = \hbar \omega$. Thus, these electrodes behave as effective half-metallic ferromagnets which emit or absorb only one spin species.

The rotating frame retarded Green function~\cite{Chen2009,Hattori2007}
\begin{equation}
\bar{\bf G}^r(E)=[E-\bar{\bf H}_{\rm rot}-\bar{\bf \Sigma}^r(E)]^{-1},
\end{equation}
and the lesser Green function
\begin{equation}
\bar{\bf G}^<(E) = \bar{\bf G}^r(E) \bar{\bf \Sigma}^<(E) \bar{\bf G}^\dagger(E),
\end{equation}
describe the density of available quantum states and how electrons occupy those states, respectively. Here $\bar{\bf H}_{\rm rot}$ is the matrix representation of $\hat{H}_{\rm rot}$~(\ref{eq:hrot}) in the basis of local orbitals. The retarded self-energy matrix $\bar{\bf \Sigma}^r(E)=\sum_{p,\sigma} \bar{\bf \Sigma}^{r,\sigma}_p(E)$ is the sum of
retarded self-energies introduced by the interaction with the leads which determine escape rates of \mbox{spin-$\sigma$} electron into the electrodes $(p,\sigma)$.

For noninteracting systems described by the Hamiltonian (\ref{eq:hrot}), the lesser self-energy is expressed in terms of $\bar{\bf \Sigma}^{r,\sigma}_p(E)$ as
\begin{equation}\label{eq:lesser_se}
\bar{\bf \Sigma}^<(E)=\sum_{p,\sigma} i f^{\sigma}(E) \bar{\bf \Gamma}_p^{\sigma}(E).
\end{equation}
The level broadening matrix in the rotating frame
\begin{equation}
\bar{\bf \Gamma}_p^\sigma(E) = -2 {\rm Im}\, \bar{\bf \Sigma}^r_p(E) = -2 {\rm Im}\, {\bf \Sigma}^r_p \left(E + s \frac{\hbar \omega}{2} \right),
\end{equation}
is obtained from the usual self-energy matrices~\cite{Haug2007} ${\bf \Sigma}^r_p(E)$ of semi-infinite leads in the laboratory frame with their energy argument being shifted by $s  \hbar \omega/2$ to take into account the ``bias voltage'' in accord with Fig.~\ref{fig:setup}(b). The distribution function of electrons in the electrodes of the rotating frame DC circuit is given by
\begin{equation}\label{eq:fermi}
f^{\sigma}(E)=\frac{1}{1+\exp[(E-E_F + s \hbar\omega/2)/kT]},
\end{equation}
where $s=+$ for \mbox{spin-$\uparrow$} and $s=-$ for \mbox{spin-$\downarrow$}.

\section{Local and total currents in two-terminal TI$|$FM$|$TI devices}\label{sec:two_terminal}

The spatial imaging of local spin currents has played an important role in understanding  how QSHE~\cite{Onoda2005a} or mesoscopic SHE~\cite{Nikoli'c2006} due to intrinsic SO couplings manifest in experimentally accessible multiterminal devices. In Fig.~\ref{fig:localpump}(a), we first establish a reference local-current-picture of pumping for \mbox{ZGNR$|$FM$|$ZGNR} device with no SO coupling in ZGNR regions. Its transport properties are governed~\cite{Zarbo2007} by the edge-localized quantum states induced by the topology of the chosen zigzag edges.~\cite{Murakami2008,Zarea2008} The non-zero SO coupling in Eq.~(\ref{eq:hlab}) converts the ZGNR regions into a 2D TI by opening energy gap and by forcing their strongly-localized states~\cite{Murakami2008,Zarbo2007} to become spin-filtered and to acquire a linear dispersion around $k_x a=\pi$ crossing the band gap~\cite{Kane2005,Zarea2008} (SO coupling additionally suppresses their unscreened Coulomb interaction~\cite{Zarea2008}).

Using the picture of such helical edge states within the TI regions of
\begin{figure}
\centerline{\psfig{file=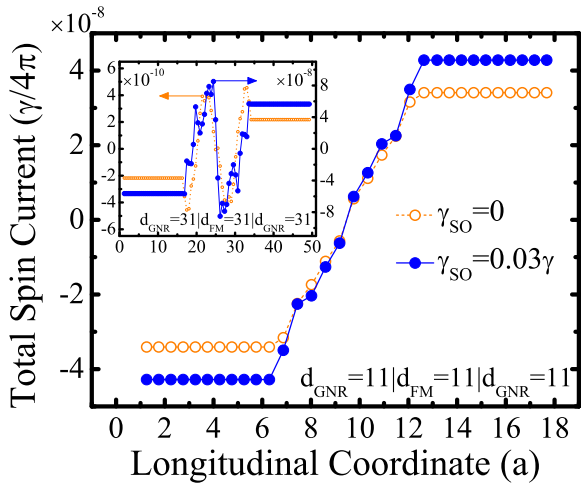,scale=0.4,angle=0}}
\caption{(Color online). Total spin current \mbox{$I^{S_z} = \sum_{ij} J_{ij}^{S_z}$} at each transverse cross section of the two-terminal GNR$|$FM$|$GNR ($\gamma_{\rm SO}=0$) and TI$|$FM$|$TI ($\gamma_{\rm SO} \neq 0$) devices  obtained by summing all bond spin currents shown in right panels of Figs.~\ref{fig:localpump}(a) and ~\ref{fig:localpump}(b), respectively. Inset shows $I^{S_z}$ for a longer device of the same width $N_z=8$. The FM magnetization is precessing with frequency \mbox{$\hbar\omega/e \approx 8.3$ $\mu$V} and cone angle \mbox{$\theta = 10^\circ$}.}
\label{fig:totalpump}
\end{figure}
the DC device in Fig.~\ref{fig:setup}(b), whose spin and chirality is illustrated in Fig.~\ref{fig:localpump}(b), we can follow possible Feynman paths of electrons in Fig.~\ref{fig:localpump}. For example, a \mbox{spin-$\downarrow$} electron from  ($L$,$\downarrow$) electrode at a higher electrochemical potential $\mu^\downarrow$ can only flow along the top left edge $\Rightarrow$ then it precesses through the FM island (because it is not an eigenstate of $\hat{\sigma}_x$ term in $\hat{H}_{\rm rot}$) $\Rightarrow$ enters with some probability into \mbox{spin-$\uparrow$} edge state on the bottom right edge $\Rightarrow$ finally, it is collected by ($R$,$\uparrow$) electrode  at a lower electrochemical potential $\mu^\uparrow < \mu^\downarrow$. The \mbox{spin-$\downarrow$} electron injected by ($R$,$\downarrow$) lead would retrace the same path in the opposite direction on the way toward ($L$,$\uparrow$) electrode.

According to these straightforward paths, one expects no chiral edge currents around bottom left and top right edges of the device. However, these currents do exist in Figs.~\ref{fig:localpump}(b) and ~\ref{fig:localpump}(c) due to more subtle backscattering effects at the \mbox{TI$|$FM} interface. They are the consequence of the paths, clearly visible in Fig.~\ref{fig:localpump}, where, e.g., \mbox{spin-$\downarrow$} electron from  ($L$,$\downarrow$) electrode is reflected and rotated~\cite{Maciejko2009} at the  \mbox{TI$|$FM} interface (where FM region breaks the time-reversal invariance) to flow backward through \mbox{spin-$\uparrow$} edge state $\Rightarrow$ then it propagates along the \mbox{left-NM-lead$|$TI} interface $\Rightarrow$ it flows along the bottom left edge, FM island, and bottom right edge to finally  enter into ($R$,$\uparrow$) electrode at electrochemical potential $\mu^\uparrow$. This process can also account for the difference between spin currents flowing along the top and bottom edges of a single TI region, which yields non-zero total spin current in Fig.~\ref{fig:totalpump}.

Another set of transverse current paths, conspicuously visible in Figs.~\ref{fig:localpump}(b) and ~\ref{fig:localpump}(c), emerges around \mbox{TI$|$FM} interfaces. These interfacial spin and charge currents are able to penetrate slightly into the bulk of the TIs,~\cite{Onoda2005a} in contrast to infinite homogeneous (i.e., all TI) systems where current flow is strictly confined to  sample edges.~\cite{Murakami2008}

Figure~\ref{fig:totalpump} plots the  total spin current along the device, which is obtained by summing local nearest-neighbor and next-nearest-neighbor bond currents at each transverse cross section in Fig.~\ref{fig:localpump}. Although non-zero locally, the total charge current through any cross section in Fig.~\ref{fig:localpump}(c), including the left and right electrodes, remains zero. Since typical spin-relaxation lengths are much longer than the length scale $\sim \hbar v_F/\Delta$ over which pumping develops,~\cite{Tserkovnyak2005} we assume that FM island is clean so that non-conserved spin currents emerge throughout its volume. The magnitude of spin current pumped into GNRs is set around the \mbox{GNR$|$FM} interface,~\cite{Chen2009} and it is {\em enhanced}  by the presence of helical edge states (for $\gamma_{\rm SO}/\gamma \lesssim 0.1$), contrary to na\"{i}ve expectation that \mbox{FM$|$TI} interface would be less transparent for spin injection.

\section{Local and total currents in six-terminal TI$|$FM$|$TI devices}\label{sec:six_terminal}

To obtain sharp conductance steps  in  quantum transport calculations of the integer QHE in  realistic (e.g., consisting of $\sim 10^5$ carbon atoms~\cite{Kazymyrenko2008}) six-terminal mesoscopic all-graphene Hall bars requires to avoid reflection at the \mbox{lead$|$sample} interface by using magnetic field both in the sample and in the semi-infinite leads, as well as by employing high quality contacts.~\cite{Kazymyrenko2008} Thus, it is somewhat surprising that perfectly quantized $G_{\rm SH}$ was obtained in Ref.~\onlinecite{Sheng2005b} for a four-terminal graphene-based 2D TI device where SO coupling is present only in the sample and whose metallic electrodes, for a chosen square lattice model, have propagating modes that couple poorly to evanescent and propagating modes within GNRs.~\cite{Robinson2007}  In fact, the corresponding contacts \mbox{metallic-electrode$|$zigzag-GNR} in those devices act as effective disorder by introducing mixing of transverse propagating modes defined by the semi-infinite ideal leads.~\cite{Areshkin2009a}

\begin{figure}
\centerline{\psfig{file=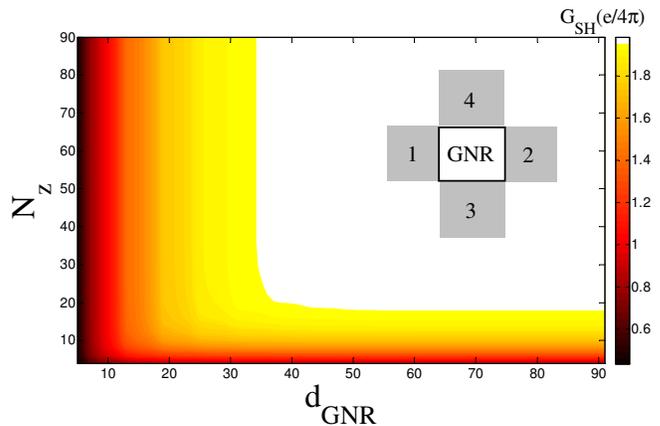,scale=0.35,angle=0}}
\caption{(Color online) The spin Hall conductance \mbox{$G_{\rm SH}=I_4^{S_z}/V$} of a voltage biased $eV_1=-eV_2=eV/2$ four-terminal bridge as a function of the size of a clean GNR central region playing the role of the QSH insulator of width  $N_z$  and length $d_{\rm GNR}$. The GNR is attached to four metallic electrodes modeled as ``lattice-matched'' square lattices whose contact to the honeycomb lattice  is illustrated in Figs.~\ref{fig:setup} and ~\ref{fig:curdef}. The transverse electrodes 3 and 4 cover the whole bottom or top zigzag edge of GNR and act as the voltage probes ($V_3=V_4=0 \rightarrow I_3=I_4=0$).}
\label{fig:size}
\end{figure}

Figure~\ref{fig:size} extends findings of Ref.~\onlinecite{Sheng2005b}, obtained for special graphene ribbon aspect ratios ($N_z=4n$, $d_{\rm GNR}={8n+1}$, $n \in \mathbb{N}$) and leads attached to a segment of the ribbon edge, to confirm that $G_{\rm SH}=2 \times e/4\pi$ can be obtained in any sufficiently~\cite{Zhou2008} wide and long graphene ribbon attached to four metallic electrodes. Here the square lattice leads cover the whole top or bottom lateral graphene edge, while the longitudinal leads are attached at armchair edges as shown in Fig.~\ref{fig:curdef}.  The square lattice of the leads is selected to be ``lattice-matched'' (lattice constant is the same as carbon-carbon distance in GNR as illustrated by Figs.~\ref{fig:setup} and ~\ref{fig:curdef}) in order to reduce  the detrimental effects of the contacts (i.e., the transmission matrix of the two-terminal device NM$|$GNR$|$NM has smaller off-diagonal elements than when ``lattice-unmatched" square lattice leads are employed to model metallic electrodes).~\cite{Areshkin2009a}

\begin{figure*}
\centerline{\psfig{file=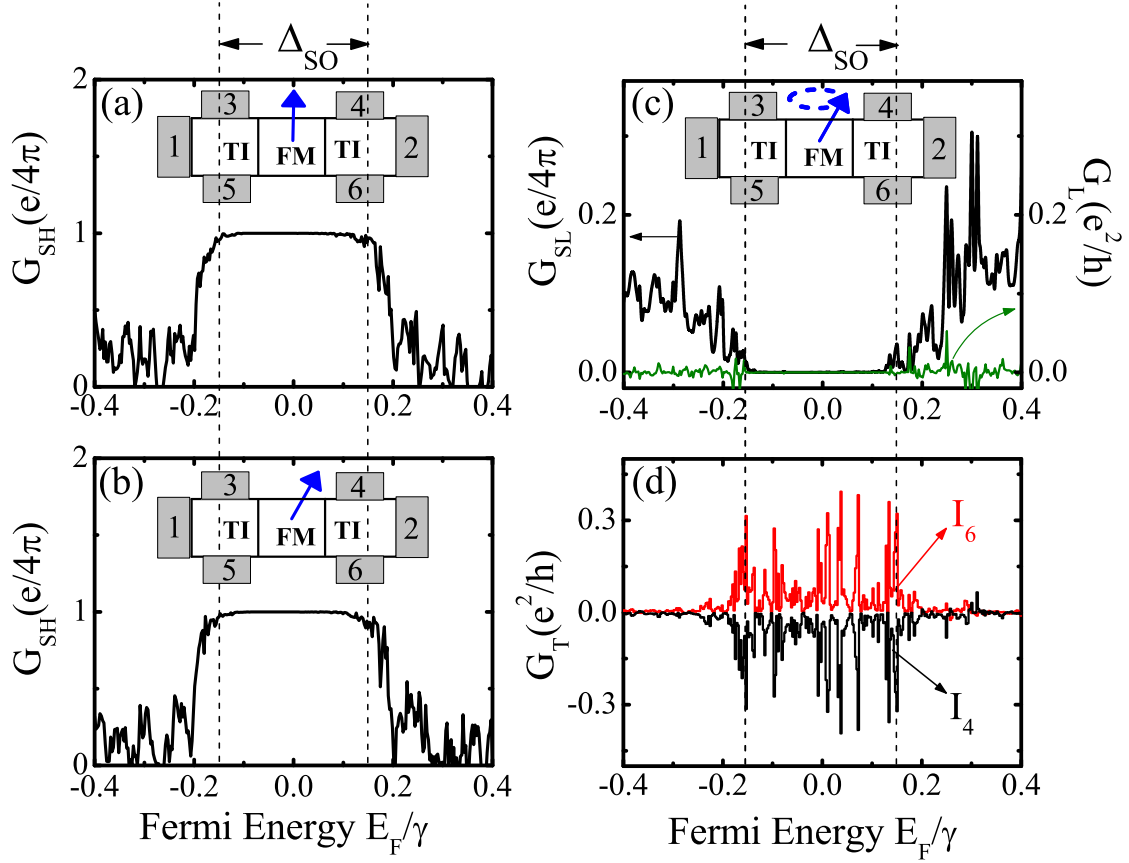,scale=0.7,angle=0}}
\caption{(Color online) (a) The spin Hall conductance $G_{\rm SH}=I_4^{S_z}/V$ of a voltage biased $V_1-V_2=V$ ($V_p=0$, {\em p}=3--6) six-terminal \mbox{$d_{\rm GNR}=65|d_{\rm FM}=32|d_{\rm GNR}=65$} device of width $N_z=32$ whose static FM magnetization is collinear with the $z$-axis orthogonal to the plane of the device. Insets label the six metallic electrodes modeled as ``lattice-matched'' square lattices shown in Fig.~\ref{fig:setup}. Panel (b) plots the same quantity $G_{\rm SH}$ when static FM magnetization is tilted by an angle $\theta = 10^\circ$ away from the $z$-axis. For the same unbiased ($V_p=0$, {\em p}=1--6) TI$|$FM$|$TI device, whose FM magnetization is precessing with frequency \mbox{$\hbar\omega/e \approx 8.3$ $\mu$V} and cone angle \mbox{$\theta = 10^\circ$}, panel (c) plots the longitudinal spin conductance $G_{\rm SL}=eI_2^{S_z}/\hbar\omega$ and longitudinal charge conductance $G_{\rm L} = eI_2/\hbar\omega$, while panel (d) shows the transverse charge conductance $G_{\rm T}=eI_4/\hbar\omega$. In panel (d) we emphasize that condition $I_4 = -I_6$,  as the signature of charge Hall current (i.e., any inverse SHE), is satisfied only within the SO energy gap $\Delta_{\rm SO}$ where GNR regions act as 2D TI. Note that the convention for the sign of the currents is $I_{p}>0$ ($<0$) and $I^{S_z}_{p}>0$ ($<0$) for particles flowing into (out of) lead $p$.}
\label{fig:sixterminal}
\end{figure*}

In order to generate  non-zero total charge current response of the 2D TIs, we attach additional four electrodes (Fig.~\ref{fig:sixterminal}) at the top and bottom edges of the device in Fig.~\ref{fig:setup}(a) of length \mbox{$d_{\rm GNR}=65|d_{\rm FM}=32|d_{\rm GNR}=65$} and width $N_z=32$. The two longitudinal electrodes and additional four transverse electrodes (covering 10 edge carbon atoms) are modeled on the same ``lattice-matched'' square lattice. Such six-terminal bridge, when biased by the voltage difference $V_1-V_2=V$, exhibits {\em quantized} $G_{\rm SH}$ shown in Fig.~\ref{fig:sixterminal}(a), despite the fact that it is not identical to standard homogeneous QSH bridges~\cite{Roth2009} since it contains FM island in the middle breaking the continuity of helical edge states. Furthermore, we confirm that quantization of $G_{\rm SH}$ is independent of the angle by which static FM magnetization is tilted from the $z$-axis, as illustrated by Fig.~\ref{fig:sixterminal}(b).

Since $\hbar\omega \ll E_F$, we can use $f^\downarrow(E)-f^\uparrow(E) \approx \hbar \omega \delta(E-E_F)$ at low temperatures for the difference of the Fermi functions present~\cite{Chen2009} in Eq.~(\ref{eq:jlab}). This ``adiabatic approximation''~\cite{Hattori2007} is analogous to linear response calculations for biased devices, allowing us to define the longitudinal spin conductance $G_{\rm SL}=eI_2^{S_z}/\hbar\omega$ and transverse charge conductance $G_{\rm T}=eI_4/\hbar\omega$ for the device whose FM magnetization is precessing with frequency $\omega$. In the unbiased ($V_p=0$) six-terminal \mbox{TI$|$FM$|$TI} device,  Fig.~\ref{fig:sixterminal}(c) shows that $G_{\rm SL}$ vanishes when $E_F$ is within the SO gap, while non-zero charge Hall currents  emerge in transverse electrodes so that $G_{\rm T} \neq 0$ in Fig.~\ref{fig:sixterminal}(d). Moreover, transverse charge currents obtained in the regime of the QSH insulator (marked by $\Delta_{\rm SO} \neq 0$ gap  in Fig.~\ref{fig:sixterminal}) are the signature of the inverse QSHE since only in this range of Fermi energies we get $I_4=-I_6$ and $I_3=-I_5$ characterizing  the usual charge Hall effect. However, we find that $G_{\rm T}=eI_4/\hbar\omega$ is not quantized. Note that the same conclusions are reached when the electrodes are made of the same TI  as in the central region of the QSH bridge.

Figure~\ref{fig:sixterminal}(c) also shows that total charge current pumped into the longitudinal leads  is zero within the QSH insulator regime ($\Delta_{\rm SO} \neq 0$), which is quite different from the conventional inverse SHE driven by spin pumping from precessing magnetization into a topologically trivial SO-coupled systems, such as the multiterminal 2DEG  with the Rashba SO coupling.~\cite{Ohe2008} In the latter case, AC charge currents (with small DC contribution vanishing as the 2DEG size increase) are pumped into both transverse and longitudinal electrodes. Their  time-dependence originates from spin non-conservation in the presence of the Rashba SO coupling, unlike in our case where charge and  \mbox{spin-$S_z$} currents are time-independent in both rotating and laboratory frames since $S_z$ spin is conserved.

\section{Discussion}\label{sec:discussion}

The presence of extended chiral edge states in QH and QSH systems makes the  analysis of transport measurements based on the Landuaer-B\"{u}ttiker multiprobe formulas~\cite{Buttiker2009} particularly simple assuming homogeneous bridge, such as the TI central region attached to electrodes made of the same TI material~\cite{Kane2005,Buttiker2009} or 2DEG and graphene in high  magnetic field attached to the same type of electrodes in the same magnetic field.~\cite{Kazymyrenko2008} In those cases,  the extended edge states are perfectly matched across the whole device and one can simply draw~\cite{Kane2005,Buttiker2009} a picture of allowed trajectories (or ``Feynman paths''), for whom the quantum edge states serve as guiding centers, to count their contribution (as ballistic one-dimensional conductors) to current in a selected electrode.

\begin{figure}
\centerline{\psfig{file=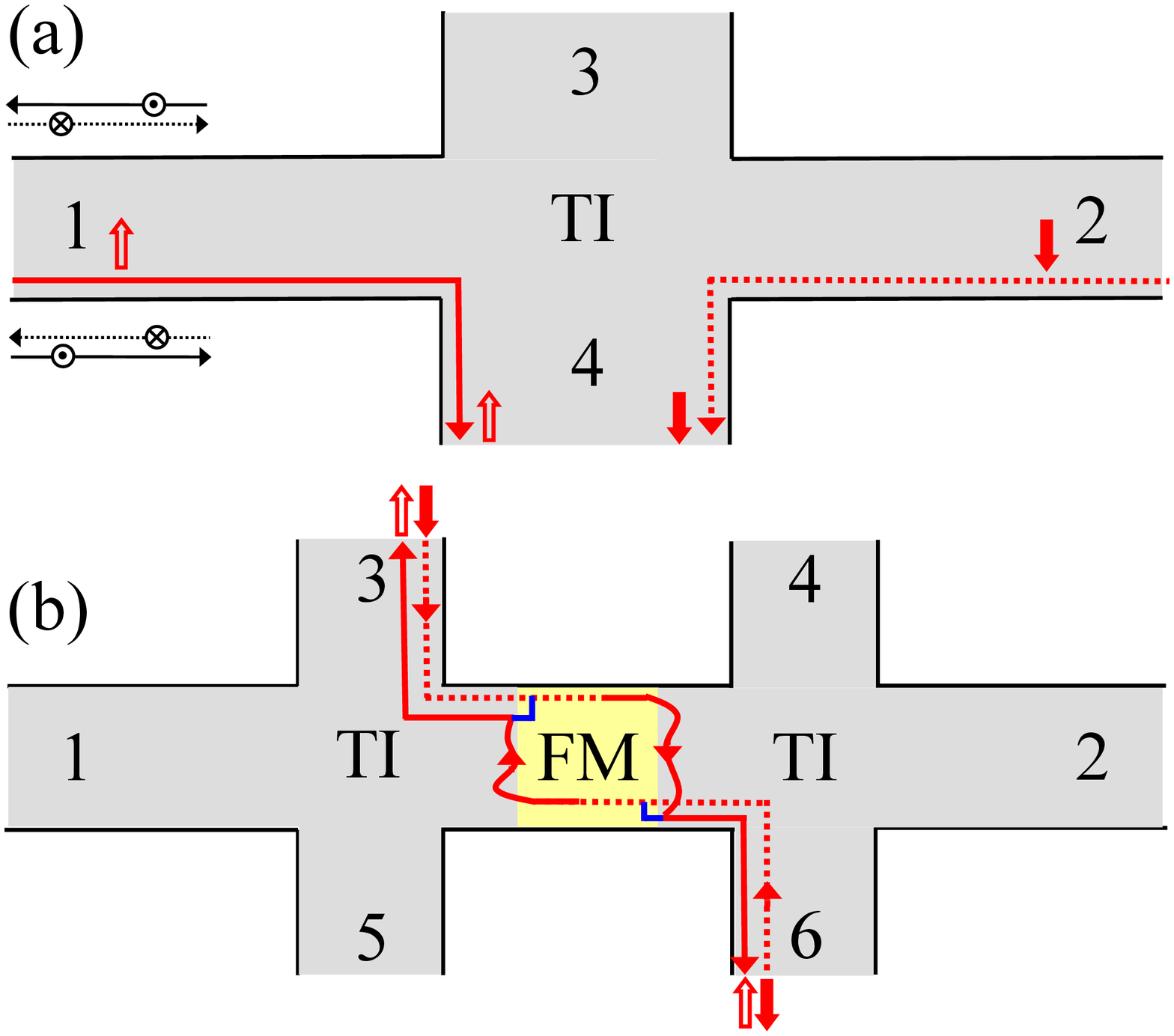,scale=0.35,angle=0}}
\caption{(Color online). Phenomenological analysis of the inverse QSHE via selected possible Feynman paths of spin-polarized injected and collected electrons for the case of: (a) four-terminal bridge of Ref.~\onlinecite{Kane2005} with homogeneous central 2D TI region (attached to TI electrodes) where pure spin current is injected from the longitudinal leads  as two counter-propagating fully spin-polarized charge currents; (b) six-terminal bridge comprised of inhomogeneous \mbox{TI$|$FM$|$TI} central region  whose  FM magnetization is precessing at frequency $\omega$ and six attached TI electrodes are made of the same material as the TI islands within the central region. The depicted Feynman paths in (b) are based on the twelve-terminal DC device in the rotating frame (with all of its electrodes accepting only one spin species and with any pair of electrodes of opposite spin being at different electrochemical potentials $\mu_p^\downarrow - \mu_{p^\prime}^\uparrow = \hbar \omega$), and they are supported by the NEGF-computed exact spatial profiles of local charge currents displayed in Fig.~\ref{fig:expl_cp}. The blue ``L-shaped'' connectors between two helical edge states of opposite spin and direction of propagation signify spin-dependent reflection~\cite{Maciejko2009} at \mbox{TI$|$FM} interfaces.}
\label{fig:expl}
\end{figure}

Following Ref.~\onlinecite{Kane2005}, we show such paths for the inverse QSHE in Fig.~\ref{fig:expl}(a) based on an abstract scheme where pure spin current is injected through the longitudinal leads due to different electrochemical potentials for \mbox{spin-$\uparrow$} and \mbox{spin-$\downarrow$} states, $\mu_1^\uparrow -  \mu_1^\downarrow = eV$ and $\mu_2^\uparrow -  \mu_2^\downarrow = - eV$ driving two counter-propagating fully spin-polarized longitudinal charge currents. In this case, continuity of the edge states and absence of any reflection between the leads and the sample ensures that conductance associated with transverse charge current is quantized.~\cite{Kane2005} However, the setup is unrealistic from the experimental viewpoint because it requires separate control of electrochemical potentials for \mbox{spin-$\uparrow$} and \mbox{spin-$\downarrow$} carriers within the same electrode.

On the other hand, the presence of FM region in our devices breaks the continuity of  edge states, thereby making the network of Feynman paths much more complicated. The analogous  issues have been explored in experiments~\cite{Haug1989} on QH bridges where the gate electrode covering small portion of the central region introduces backscattering between spatially separated edge states, thereby requiring complicated trajectory network~\cite{Haug1989} or spatial profiles of local currents~\cite{Gagel1996} to explain non-quantized features in the longitudinal resistance. Furthermore, in the case of our six-terminal \mbox{TI$|$FM$|$TI} bridge, the breaking of time-reversal invariance at the \mbox{TI$|$FM} interface by the nearby FM island also enables spin-dependent reflection where incoming electron from a helical edge state has its spin rotated to be injected in the counter-propagating helical edge state along the same edge of the sample.~\cite{Maciejko2009}

\begin{figure*}
\centerline{\psfig{file=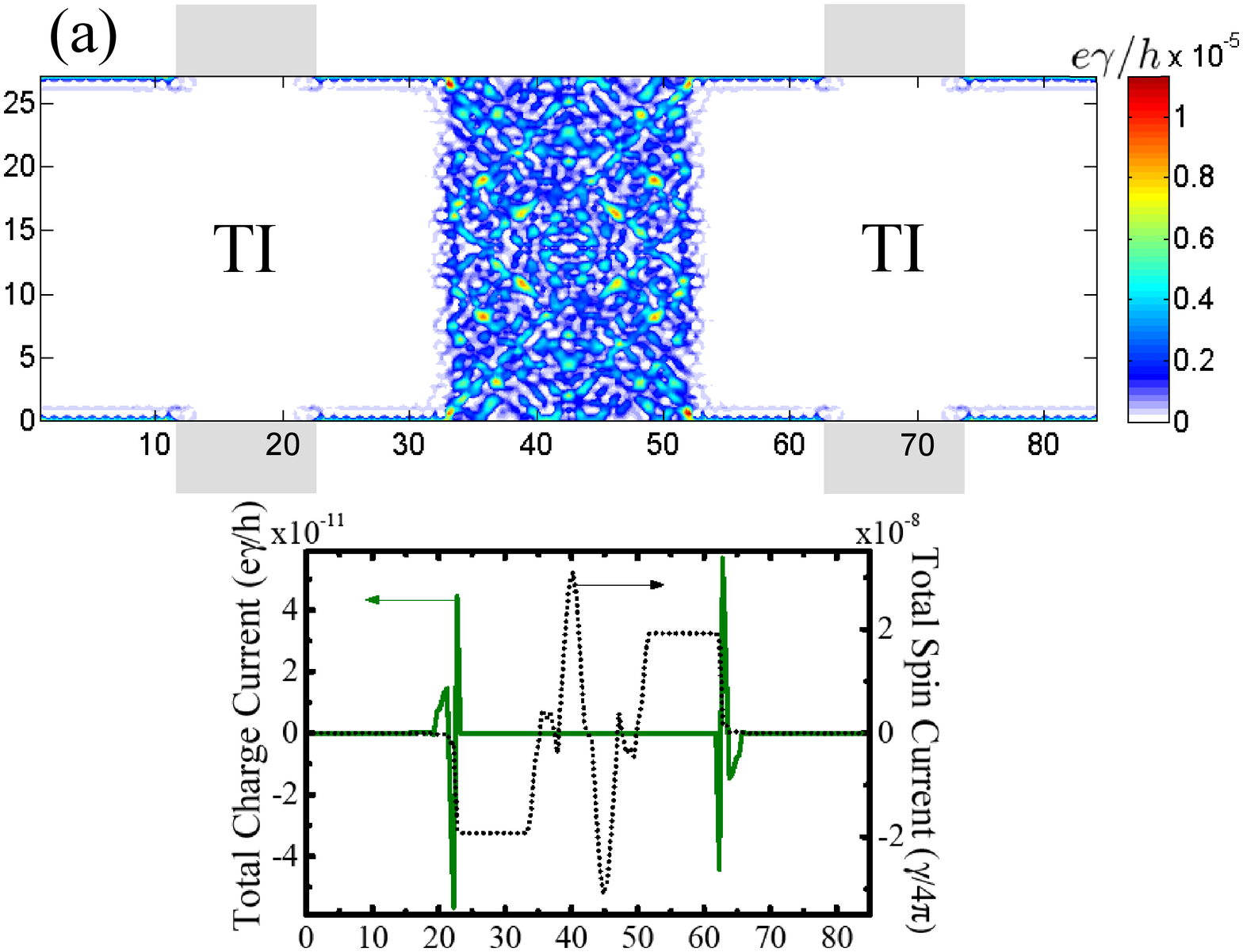,scale=0.45,angle=0}}
\vspace{0.3in}
\centerline{\psfig{file=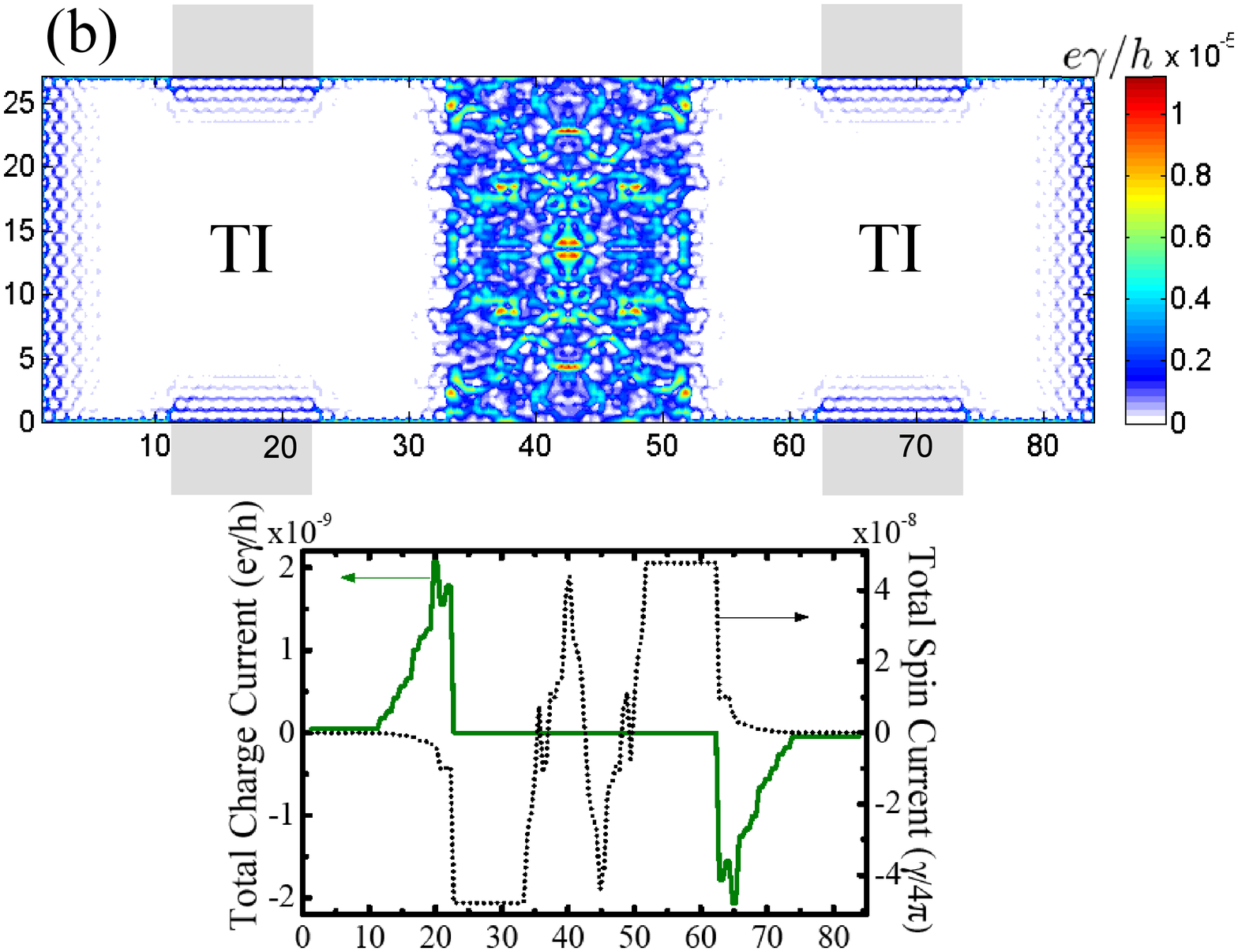,scale=0.45,angle=0}}
\caption{(Color online) Spatial profile of charge currents in six-terminal devices whose central \mbox{TI$|$FM$|$TI region} with precessing FM magnetization is attached to: (a) six TI electrodes identical to central TI islands [as assumed also in Fig.~\ref{fig:expl}(b)]; (b) six NM electrodes (as in Fig.~\ref{fig:sixterminal}) modeled on the ``lattice-matched'' square lattice. The possible Feynman paths corresponding to current profiles in (a), as well as the labeling of the six terminals, is shown in Fig.~\ref{fig:expl}(b).  The inset below each spatial profile depicts  the corresponding total spin $I^{S_z} = \sum_{ij} J_{ij}^{S_z}$ and charge currents $I^{S_z} = \sum_{ij} J_{ij}$ computed at each transverse cross section of the device as one moves from the left to the right longitudinal electrode.}
\label{fig:expl_cp}
\end{figure*}

Nevertheless, some of the important paths can be extracted from spatial profiles of local currents in Fig.~\ref{fig:localpump} for the two-terminal \mbox{TI$|$FM$|$TI} device or Fig.~\ref{fig:expl_cp} for the six-terminal \mbox{TI$|$FM$|$TI} device. This allows us to explain all of the key results on the inverse QSHE driven by spin pumping shown in Fig.~\ref{fig:sixterminal} for total terminal currents. The possible electron paths are easier to draw for  the six-terminal \mbox{TI$|$FM$|$TI} device whose electrodes are made of the same TI, and can be understood using the picture of twelve effectively half-metallic FM electrodes in the rotating frame which try to inject their fully spin-polarized electrons into  chiral spin-filtered edge states moving in proper direction.

For example, Fig.~\ref{fig:expl}(b) shows \mbox{spin-$\downarrow$} electrons starting from lead (3, $\downarrow$) at a higher electrochemical potential $\mu_3^\downarrow$  to enter the right moving \mbox{spin-$\downarrow$} helical state on the top edge. The electrons have probability to penetrate into the FM island where they precess and continue to propagate down (e.g., through interfacial spin currents shown in Fig.~\ref{fig:localpump} and  Fig.~\ref{fig:expl_cp} along the right \mbox{FM$|$TI} interface)  to finally enter the bottom helical edge state as \mbox{spin-$\uparrow$} particles which allows them to be collected by the electrode \mbox{(6, $\uparrow$)} at lower electrochemical potential   $\mu_6^\downarrow - \mu_3^\uparrow = -\hbar\omega$. At the same time, incoming electrons from lead (3, $\downarrow$) can also be reflected at the \mbox{TI$|$FM} interface where accompanying spin rotation~\cite{Maciejko2009} makes it possible for them to propagate back into the lead (3,$\uparrow$) which they enter as \mbox{spin-$\uparrow$} electron with some probability. Portions of this path are retraced by \mbox{spin-$\downarrow$} electron from lead (6,$\downarrow$) flowing toward lead (3,$\uparrow$), where reflection and spin rotation at the right \mbox{TI$|$FM} interface generate additional current of \mbox{spin-$\uparrow$} electrons into  lead 6 to produce total charge current $I_6 > 0$ whose sign is compatible with the exact calculations shown in Fig.~\ref{fig:sixterminal}(d). These paths also explain the absence of quantization of $G_{\rm T}$ conductance in Fig.~\ref{fig:sixterminal}(d) since transport which generates non-zero longitudinal charge currents is not confined solely to the helical edge states as in Fig.~\ref{fig:expl}(a).

While the selected paths in Fig.~\ref{fig:expl} are compatible with complete  (numerically exact) spatial profiles of local charge currents shown in Fig.~\ref{fig:expl_cp}(a), the profiles  also contain trivial strictly edge paths carrying non-zero local charge and spin currents between leads 4 and 2 due to ballistic transport of \mbox{spin-$\downarrow$} electrons from electrode $(4,\downarrow)$ toward  $(2,\downarrow)$ with higher state occupancy then for \mbox{spin-$\uparrow$} electrons propagating from $(2,\uparrow)$ toward  $(4,\uparrow)$ electrode. Since the same local currents exists along the opposite edge connecting electrodes 6 and 2, the {\em total} spin or charge current flowing into the longitudinal electrode 2 is identically equal to zero [see inset below Fig.~\ref{fig:expl_cp}(a)]. The same conclusion holds by symmetry for leads 1, 3, and 5 on the opposite side of bridge. Thus, the absence of any Feynman paths that would allow electrons from (1,$\downarrow$) and (2,$\downarrow$) electrodes to reach some other ($p$,$\uparrow$) electrode via propagation through helical edge states, combined with trajectories within FM or along the \mbox{TI$|$FM} interface, confirms one of the key results in Fig.~\ref{fig:sixterminal}(c)---absence of longitudinal spin and charge currents in the spin-pumping-induced inverse QSHE.

We also note that the above discussion becomes  more complicated for the \mbox{TI$|$FM$|$TI} bridges with six NM electrodes that were actually employed in Fig.~\ref{fig:sixterminal}. This is exemplified by Fig.~\ref{fig:expl_cp}(b) where new paths have to be take into account that allow electrons to propagate along interfaces between TI regions and the attached NM electrodes while also penetrating~\cite{Onoda2005a} into the bulk of NM electrodes where there are no helical edge states. In the case of leads 1 and 2, local currents around \mbox{TI$|$NM} interface penetrate only slightly into the bulk of the NM electrodes  and then return toward TI to ensure zero total spin and charge currents in the longitudinal electrodes, as demonstrated by the inset below Fig.~\ref{fig:expl_cp}(b).

\section{Concluding Remarks}\label{sec:conclusions}

Using the  mapping of time-dependent spin pumping by precessing magnetization to a multiterminal DC device in the rotating frame, we describe injection of thus generated local and total spin currents into helical edge states of a topological insulator in the absence of any externally applied bias voltage. In the regime where the voltage biased six-terminal \mbox{TI$|$FM$|$TI} nanodevice exhibits QSHE, the corresponding unbiased device with precessing FM magnetization generates  charge currents in the transverse electrodes (characterized by $I_4=-I_6$ and $I_3=-I_5$), while bringing pumped total spin and charge currents in longitudinal electrodes to zero. Although the transverse charge conductance of such inverse QSHE is not quantized, these {\em two responses} to pumping can be used to probe the TI phase via unambiguous {\em electrical} measurements. The absence of quantization of transverse charge conductances was explained as the consequence  of  electron propagation paths being composed of both simple segments guided by chiral spin-filtered  edge states of finite extent and more complicated segments through the FM island or  around the \mbox{TI$|$FM} interface where spin-dependent reflection accompanied by spin rotation can take place. Our analysis based on imaging of local (i.e., on the scale of the lattice constant) spin and charge transport reveals  interfacial currents around \mbox{TI$|$FM} interfaces that can penetrate slightly into the bulk of the TI or interfacial currents around \mbox{TI$|$NM-electrode} contacts which break the continuity of helical edge states.

Although a topologically trivial system with the intrinsic SO couplings (such as the Rashba spin-split 2DEG~\cite{Nikoli'c2006}) would also generate non-quantized transverse charge currents in response to pure spin currents pumped by precessing magnetization,~\cite{Ohe2008} this setup also pumps longitudinal charge currents.~\cite{Ohe2008} This is completely different from the behavior of the proposed \mbox{TI$|$FM$|$TI} multiterminal device. In addition, for SO couplings which do not conserve $S_z$ spin (such as the Rashba one), both transverse and longitudinal pumped charge currents are time-dependent,~\cite{Ohe2008} as opposed to DC transverse charge Hall currents generated by our device.

During the preparation of this manuscript we became aware of the theoretical proposal~\cite{Qi2008} for a charge pumping device designed to induce electrical response of a 2D TI where two FM islands, one with precessing and one with fixed magnetization, are deposited on the top and bottom edges, respectively, of a QSH insulator attached to two electrodes. While the physical motivation  leading to this device---fractional charge response to magnetic domain wall acting as external physical field---is different from ours, its operation can be easily explained from our four-terminal DC device picture (Fig.~\ref{fig:setup}) of spin and charge pumping in the rotating frame. That is, the top FM island (see Fig.~3 in Ref.~\onlinecite{Qi2008}) injects pure spin current into  upper helical edge states of the QSH insulator, which is then partially converted into charge current flowing along the edges as explained by spatial profiles of local charge currents in Fig.~\ref{fig:localpump}(c). The role of the bottom FM island in this setup is to block the local charge current, which otherwise flows in the opposite direction along the bottom edge in Fig.~\ref{fig:localpump}(c) leading to $I_1=I_2=0$ in our device setup. Such blocking is based on a simple observation that FM region whose magnetization is collinear with the magnetization of the effective half-metallic electrodes in the rotating frame does not permit any transport through it that connects these electrodes. However, it is unclear if such two-terminal device can display perfectly quantized longitudinal charge current $I= e \omega / 2 \pi$ (or, equivalently, quantized conductance $G_{\rm L}=eI_2/\hbar\omega = e^2/h$ in our notation) in a realistic setup, which was conjectured in Ref.~\onlinecite{Qi2008} via qualitative arguments without treating the effect of \mbox{TI$|$FM} interfaces on adiabatic charge pumping through either the  scattering~\cite{Tserkovnyak2005} or NEGF~\cite{Chen2009,Hattori2007} approaches.

\begin{acknowledgments}
This work was supported by DOE Grant No. \mbox{DE-FG02-07ER46374} through the Center for Spintronics and Biodetection at the University of Delaware. C.-R. Chang was supported by the Republic of China National Science Council Grant No. \mbox{95-2112-M-002-044-MY3}.
\end{acknowledgments}





\end{document}